# ACWA: An AI-driven Cyber-Physical Testbed for Intelligent Water Systems

Feras A. Batarseh[1, 2, 3]^, Ajay Kulkarni[2], Chhayly Sreng[3], Justice Lin[3], and Siam Maksud[1]

[1] Department of Biological Systems Engineering, Virginia Tech, Blacksburg, VA 24060, USA
(E-mail: batarseh@vt.edu, siam@vt.edu)

[2] Commonwealth Cyber Initiative, Virginia Tech, Arlington, 22203, USA
(E-mail: ajaysk@vt.edu)

[3] Department of Electrical and Computer Engineering, Virginia Tech, Arlington, 22203, USA
(E-mail:chhaylysreng@vt.edu, justicelin5403@vt.edu)

^Corresponding author

## Abstract

This manuscript presents a novel state-of-the-art cyber-physical water testbed, namely: The AI and Cyber for Water and Agriculture testbed (ACWA). ACWA is motivated by the need to advance water resources' management using AI and Cybersecurity experimentation. The main goal of ACWA is to address pressing challenges in the water and agricultural domains by utilising cutting-edge AI and data-driven technologies. These challenges include Cyberbiosecurity, resources' management, access to water, sustainability, and data-driven decision-making, among others. To address such issues, ACWA consists of multiple topologies, sensors, computational nodes, pumps, tanks, smart water devices, as well as databases and AI models that control the system. Moreover, we present ACWA simulator, which is a software-based water digital twin. The simulator runs on fluid and constituent transport principles that produce theoretical time series of a water distribution system. This creates a good validation point for comparing the theoretical approach with real-life results via the physical ACWA testbed. ACWA data are available to AI and water domain researchers and are hosted in an online public repository. In this paper, the system is introduced in detail and compared with existing water testbeds; additionally, example use-cases are described along with novel outcomes such as datasets, software, and AI-related scenarios.

## Highlights

a. ACWA is a novel cyber-physical testbed for testing, validating, and experimenting with intelligent water systems and AI.

b. We compare against all existing water testbeds, and illustrate the benefits of leveraging ACWA's design and data.

c. The testbed consists of multiple sensors, computational nodes, pumps, pipes, valves, tanks, and other computational components.

d. ACWA is the first data generator of its kind (hardware and software) in the water domain.

e. Open water domain datasets with multiple variables are available, including ones that enable the testing of Cyberbiosecurity and AI deployments in water systems.

## 1. Introduction

Water Supply Systems (WSS) are primary sources that provide and ensure water quality and availability, a prerequisite for flourishing economies, protecting public health, and promoting national prosperity (Batarseh & Kulkarni, 2023). However, the lack of water availability has become a growing concern that requires increasing attention. The United Nations (UN) World Water Development Report of 2021 (2021) states that the global population is experiencing severe water scarcity and will rise from 32 million people in 1900 to 3.1 billion people by 2050.

Additionally, cyber threats on WSS have been a critical threat to water resources in the United States and around the world. In WSS, it is common to use Supervisory Control and Data Acquisition (SCADA), a standard "legacy" software infrastructure, making systems susceptible to cyber threats (CISA, 2021). Recently, Hassanzadeh et al. (2020) reported fifteen cyber incidents on WSS and mentioned that more sophisticated attacks, such as data poisoning, minimum perturbations, and botnets, require algorithms that can detect, classify, and counter adversarial actions. According to Batarseh and Kulkarni (2023), AI is the leading approach to such defences due to its ability to identify unwarranted pattern shifts in networks and datasets. Despite AI being the primary approach, there is an absence of historical data (Sobien et al., 2023) concerning cybersecurity measures in water resources management. The dissemination of water data is rare due to their lack of integration and interoperability across various data archives (Bustamante et al., 2021). Further, difficulties in data accessibility arise from challenges in locating, obtaining, and comparing data across different regions, given that various parties and agencies store and manage data in different formats and have different governance limitations. Additionally, there are inconveniences in data accessibility caused by political, economic, and cultural barriers, along with varying legal requirements. For instance, most WSS have restrictions that stop them from sharing their data, also especially with cyber-related data, there is a challenging scarcity in data points that could be used for identifying patterns via AI and other methods. Considering these challenges and their severity, the need to create an innovative testbed that can create data and validate scenarios in a dynamic manner is evident. ACWA therefore, is an innovative cyber-physical system designed to tackle these challenges by generating valuable scenarios and datasets through experimental procedures. The underlying concept of the ACWA Lab involves creating a water testbed with varying topologies (i.e., dynamic structures) to simulate diverse water distribution scenarios. Moreover, a soil topology is also established to explore the impacts of water on irrigation and agricultural practices. This comprehensive setup will incorporate different sensors (water and soil) and chemical agents to enable seamless data collection, feeding directly into a database. The accumulated data within the database is visualised via real-time dashboards, offering proactive monitoring and evaluation capabilities.

## 2. Related Work

A testbed can be defined as a "composite abstraction of systems used to study system components and interactions to gain further insight into the essence of the real system" (Fortier & Michel, 2003). This section reviews seven existing testbeds developed in the water domain for simulation of monitoring and control processes. The reviewed testbeds are shown in Table 1, the testbeds are also compared and contrasted with ACWA.

Kartakis et al. (2015) presented WaterBox; a small-scale testbed that allows the simulation of monitoring and control processes of smart water systems. The WaterBox is a closed-loop structure consisting of three individual layers, as shown in Figure 1. The upper layer is the Supply, which simulates the reservoir and a pumping station. The WaterBox also has pressure sensors and controllable values to monitor the water transfer to the middle layer. The middle layer indicates three District Metered Areas (DMAs) represented by tanks of different sizes that provide water to the lower layer. The lower layer represents Demand, which mimics the water demand variation in time using valves. Each output from the DMA in the lower layer has a flow sensor and valve installed. In the end, the water from the lower layer is collected in a large tank and recycled to the reservoir using an underwater pump. The details on information on preliminary experiments using WaterBox are provided by Kartakis et al. (2015).

*Table 1: Seven existing testbeds developed in the water domain*

| Water testbed | Main declared purpose |
|---|---|
| WaterBox (Kartakis et al., 2015) | Simulation of monitoring and control processes of smart water networks |
| EARNPIPE (Karray et al., 2016) | Detection of leaks and localization |
| Secure Water Treatment (SWaT) (Goh et al., 2017) | Representation and simulation of scaled-down version of real-world water treatment |
| Failure and Attack on interdependent Critical InfrastructurES (FACIES) (Bernieri et al., 2017) | Emulate a water system of a small city |
| PLC-based water system (Laso et al., 2017) | Detection of anomalies and malicious acts in cyber-physical systems |
| Water Distribution testbed (WADI) (Ahmed, 2017) *and* the Security Water Processing (SWaP) testbed (Calder et al., 2023) | Representation and simulation of scaled-down version of a large water distribution network in a city - SWaP, an add-on, is an industrial control system for security research and training |
| Smart water campus (Oberascher et al., 2022) | Water systems for fault detection and utilisation of cross-system rainwater harvesting |

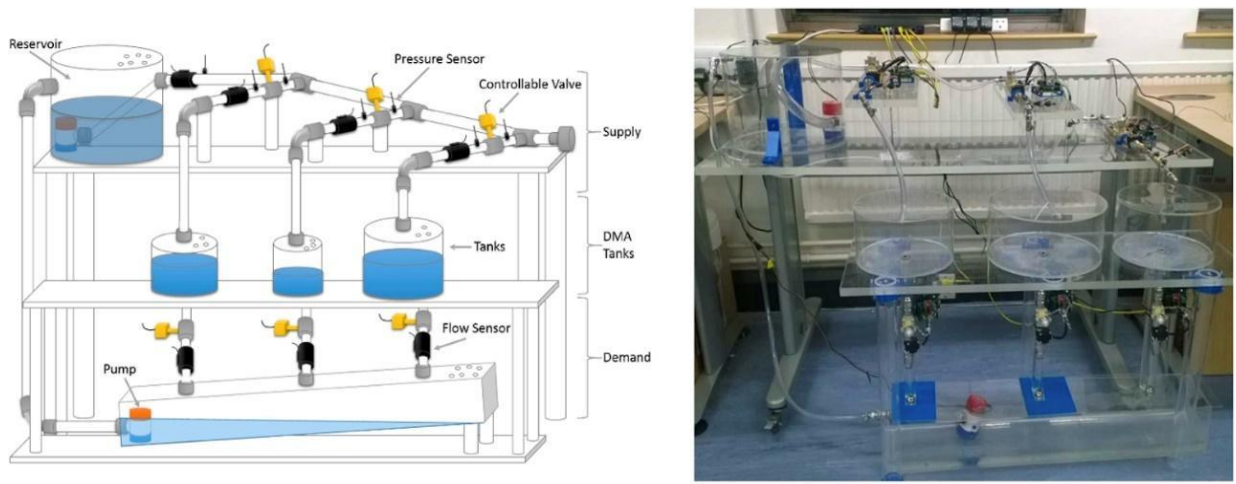

*Figure 1: WaterBox testbed is a closed-loop structure consisting of three individual layers (Kartakis et al., 2015)*

Karray et al. (2016) presented EARNPIPE to provide a low-power Wireless Sensor Network (WSN) solution that detects leaks (and provides localization). The meaning of localization in this context is the sectional leaks in an otherwise leakage-free pipe. As Figure 2 indicates, this testbed is constructed in a rectangular shape with polyethylene pipes, which supports pressure up to 12 bar (1 bar = 100,000 pa). Further, two valves are installed as an inlet and outlet to vary water demand by varying the pressure. EARNPIPE also includes a 1000 m3 reservoir for water storage and an electric pump with one Horsepower (hp) for pumping water from the reservoir to pipes. To induce leaks, two garden taps are used in their setup. The authors also have proposed and discussed the implementation of leak detection (Leak Detection Predictive Kalman Filter) and localization algorithms (Modified Time Difference of Arrival) (Karray et al., 2016).

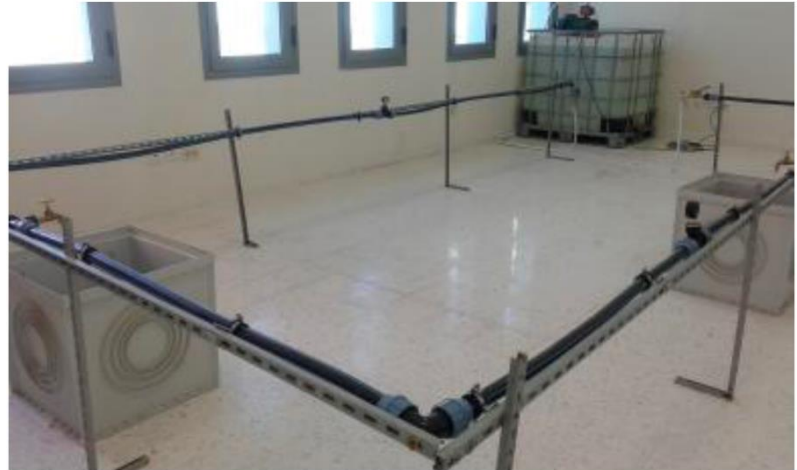

*Figure 2: EARNPIPE testbed is developed with polyethylene pipes (Karray et al., 2016)*

Goh et al. (2017) presented a six-stage Secure Water Treatment (SWaT) testbed, a scaled-down version of real-world water treatment plant. SWaT utilises membrane-based ultrafiltration and reverse osmosis units to produce 5 gallons per minute of filtered water. The overall process layout and the architecture of SWaT are presented in Figure 3. SWaT has six main processes: taking raw water (P1), pre-treatment (P2), filtration via membranes (P3), dichlorination (P4), reverse osmosis (P5), and distribution (P6). The authors collect data from these processes. Network traffic data are also collected from commercial equipment via Check Point Software Technologies Ltd. The additional details of data generation and attacks simulation using the SWaT testbed are discussed inGoh et al. (2017).

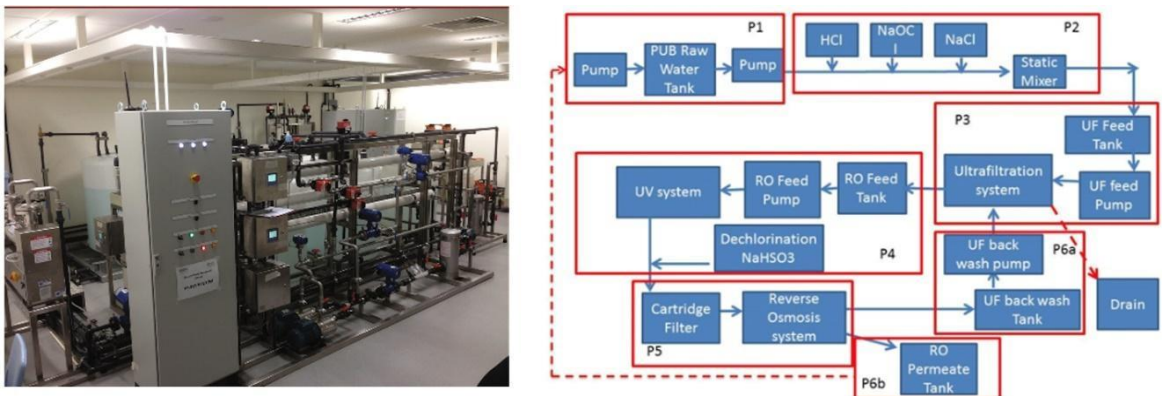

*Figure 3: Overall process layout and architecture of SWaT (Goh et al., 2017)*

Bernieri et al. (2017) presented the Failure and Attack on Interdependent Critical InfrastructurES (FACIES) testbed developed within an European Union (EU) project. The system emulates a water system of a small city. FACIES testbed consists of 5 tanks connected by pipes where each tank consists of a water level sensor. The water flow in the system is managed by four centrifugal pumps, 20 solenoid valves, and seven manual valves. The authors noted that these tanks could be arranged in 14 different configurations to perform a variety of experiments. The FACIES testbed and other modules present in FACIES are shown in Figure 4. The real-time data from the water level sensors is collected by Modicon M340 and Schneider Electric PLCs, which are stored on a local database developed in Oracle MySQL Workbench. Additional details on FACIES can be found in Bernieri et al. (2017).

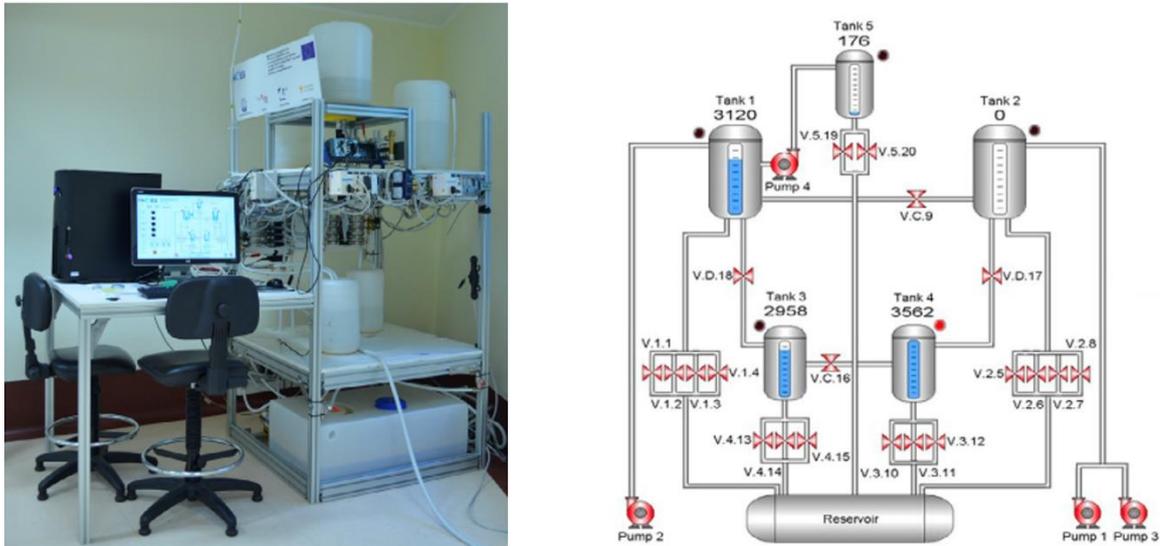

*Figure 4: The FACIES testbed and the modules constituting the testbed (Brenieri et al., 2017)*

Laso et al. (2017) presented a dataset generated from the physical water testbed to enable the detection of anomalies and malicious acts in cyber-physical systems. This testbed utilises two tanks (one with a 7-Litre capacity and the other with a 9-Litre) for storing water or fuel, one ultrasound depth sensor, four discrete sensors, and two pumps. As shown in Figure 5, the physical components are controlled and monitored using a computer connected to a Programmable Logic Controller (PLC). This water testbed simulates 15 unique situations affecting ultrasound sensors, discrete sensors, the underlying network, or the whole subsystem.

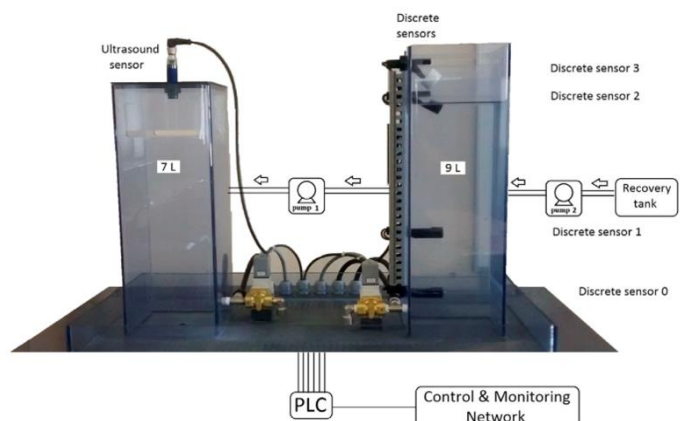

*Figure 5: A physical water testbed used by Laso et al. (2017)*

Ahmed et al. (2017) presented an architecture of a Water Distribution testbed (WADI), a scaled-down version of a large water distribution network of a city, in 2023, the same team released an additional SWaP testbed as indicated in Table 1. WADI consists of three stages (P1, P2, and P3), as indicated in Figure 6, and when in operation, it can supply 10 gallons/min of filtered water. The P1 is a primary grid that contains two water tanks (2500 Litres each), a water level sensor, and a chemical system that maintains the water quality. Additionally, water quality sensors are also installed in P1. The second stage is P2, which indicates the secondary grid. This stage consists of two elevated reservoir tanks and six consumer tanks. The raw tanks from P1 supply water to these elevated tanks based on pre-set demand patterns set by the authors. Lastly, P3 deals with the return water grid, which is also equipped with a tank. When the demands of the consumer tanks are met in P2, the water drains to the return water grid. Detailed information on communication infrastructures and how they can help analyse an attack's impact is provided by Ahmed et al. (2017).

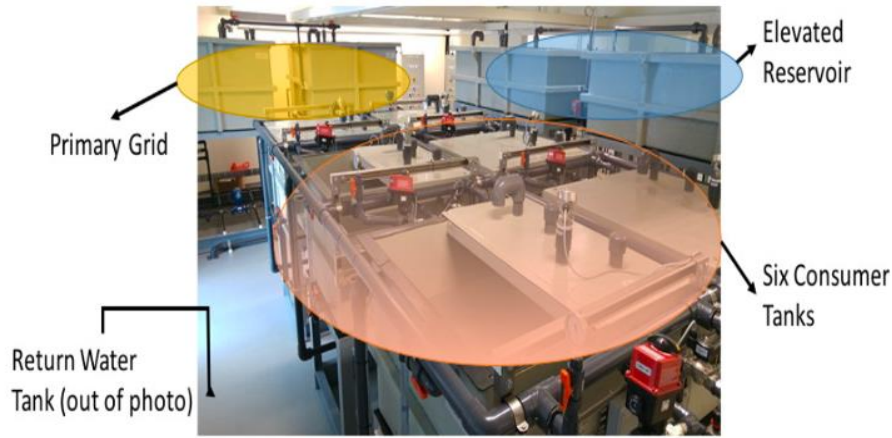

*Figure 6: Primary grid (P1), elevated reservoir and six consumer tanks (P2) and return water grid (P3) in WADI (Ahmed et al., 2017)*

Oberascher et al. (2022) described the Smart water campus testbed for monitoring water networks that can be useful for fault detection in real-time along with involving cross-system improvements like rainwater harvesting. The authors noted that the Smart water campus is a network-based urban water infrastructure that integrates a water distribution network, an urban drainage network, and nature-based solutions. The testbed leverages perception, communication, middleware, and processing layers. An illustration of the Smart water campus testbed is provided in Figure 7. A detailed description of the four layers used in the Smart water campus testbed and a demonstration of smart applications can be found in Oberascher et al. (2022).

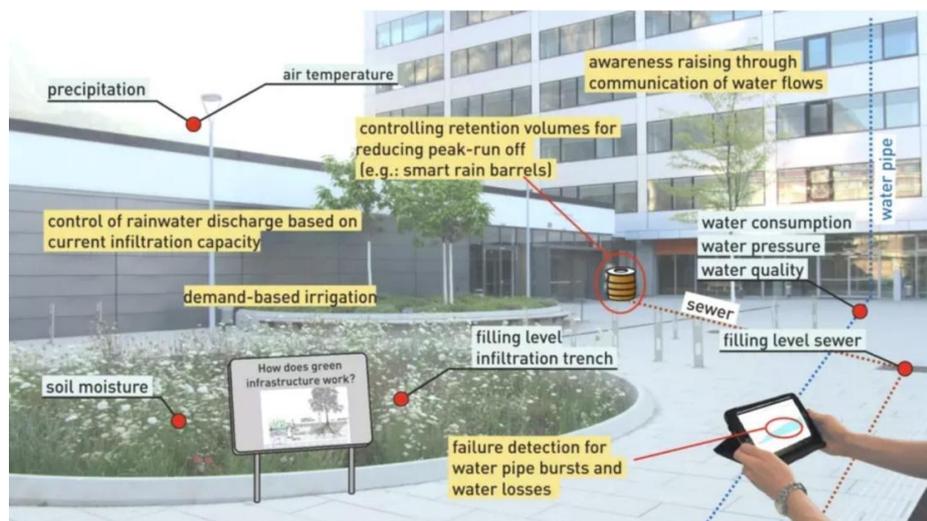

*Figure 7: Illustration of features of the Smart water campus (Oberascher et al., 2022)*

The scope of most of the existing testbeds summarised in this section is either within one or two experimental spectrums. For example, WaterBox (Kartakis et al., 2015) focuses on the simulation of monitoring and control processes of water networks and EARNPIPE (Karray et al., 2016) is designed to detect leaks and localization, ACWA (the new testbed presented in this paper) aims to provide a dynamic environment that allows for multiple scenarios, including related to irrigation and soil, a notion missing from existing testbeds. SWaT (Goh et al., 2017), WADI (Ahmed, 2017), and FACIES (Bernieri et al., 2017) are explicitly developed for studying the effects of attacks (cyber and physical) on water plants while Laso et al. (2017) presented a testbed for anomaly/malicious act detection. These testbeds involve components such as pipes, pumps, and valves along with flow and water level sensors, but these testbeds do not capture water chemicals and related parameters (which is commonly a manner of attack on water systems), the lack of such experimentation renders these testbeds limited. Smart water campus (Oberascher et al., 2022) is the only testbed focusing on water quality parameters and on capturing soil and climate-related variables, although it's spread across eight hectares and doesn't provide the ability to run distributed AI algorithms. Considering these aspects, the ACWA testbed differentiates from the reviewed testbeds via the following main characteristics:

1. Modularity: The components of the testbed (multiple water networks) are designed modularly. This unique flexibility allows for the creation of new topologies and to add more nodes (such as: pumps, valves, pipes, tanks, sensors, soil, reservoirs) in any topology based on the experiment's design and goals.
2. Sensors: The ACWA testbed has fourteen water sensors that capture data on traditional water parameters, such as: water level, water pressure, and water flow; and quality variables, such as: pH, temperature, Dissolved Oxygen (DO), nitrate, and Electrical Conductivity (EC) - among other data points listed in this manuscript. Additionally, the soil topology is equipped with a soil moisture sensor, water turbidity, and two soil probes to capture EC and temperature data.
3. Data: The big data created at ACWA are captured at 1, 5, and 30-seconds intervals and are stored in a local MongoDB database. This unique feature allows researchers to capture granular-level data and analyse miniscule changes in sensor values, including the deployment of AI and security algorithms. A data collection frequency not found anywhere else.
4. Water and soil: The ACWA testbed is carefully designed and planned to accommodate a soil topology with four soil beds along with Line, Star, and Bus water network topologies.

Furthermore, there is a lack of existing software water simulators, the most commonplace water simulator is EPANET, a prolonged hydraulic and water quality dynamics simulator built within a pressurised piping systems environment; based on Rossman (2000), EPANET is also used in commercial software development – in our open access GitHub repository (https://github.com/AI-VTRC/ACWA-Data), we also share the EPANET model for ACWA (in Appendix E), accessible to other researchers. A main addition that we present via ACWA is also a digital twin, namely: the ACWA simulator, also presented in this paper. The next section presents the main parts of ACWA, the main contribution of this paper.

## 3. Methods

To dynamically simulate and collect real-time data on different scenarios of WSS, the lab's sensors are built into four topologies to capture water and soil quality attributes such: as pH, temperature, DO, turbidity, nitrate levels, EC, soil moisture, water level, pressure, and flow rate. These data are stored in a MongoDB database for analysis, model development, and AI assurance, which can assist in tackling rising challenges in the water and agricultural domains as found in recent literature (Batarseh & Kulkarni, 2023; Saad et al., 2020; Dobermann et al. 2004; Li et al., 2020).

### 3.1 The Cyber-Physical Water Topologies

A topology is a physical and logical arrangement of nodes and connections in a network. Literature (Liu & Liu, 2014; Peterson & Davie, 2007) indicates seven commonly used network topologies: Line, Bus, Star, Ring, Mesh, Tree, and Hybrid. These topologies are foundational in computer networks. However, literature on WSS indicates four main structures: Grid-Iron, Ring, Radial, and Dead-end (Adeosun, 2014). These four structures can be designed using the fundamental computer network topologies. In Dead-end WSS (National Research Council, 2007), the main line is at the centre, and sub-main lines are divided into branches. This structure can be developed using a Bus topology at ACWA in which the nodes are connected to a central line using a branch-like structure. Similarly, Ring WSS is a circular system, and a line topology can be turned into a Ring WSS if we connect the first and last nodes of a network. Also, in Radial WSS (National Research Council, 2007), a reservoir is present at the centre, and water is distributed directly from it, which can be represented using ACWA's Star topology. Considering these aspects and motivation from computer networks, we have built Line, Star, and Bus topologies. In these topologies, nodes are represented by water tanks and they are connected to pumps and reservoirs using pipes and tubes. These topologies are placed independently on three tables of sizes 5 inches x 2.6 inches x 2.5 inches (length x width x height), which can also be connected into a hybrid topology using Poly Vinyl Chloride (PVC) or Chlorinated Poly Vinyl Chloride (CPVC) pipes. In our setup, we used two big water tanks (35 Gallons) as reservoirs to provide water to the testbed. The detailed descriptions of these three topologies are as follows.

### 3.1.1. Line Topology
The Line topology is developed by building point-to-point connections between three tanks with pumps, as shown in Figure 8a. To build this topology, three rectangular tanks and two diaphragm water pumps are connected via ½ inch PVC pipes. In Line topology, one diaphragm water pump pumps the water from the reservoir to three tanks, and the other pump pulls the water out from the tanks back to the reservoir. For every tank, the water flow can be changed at any point using manual valves installed in the network. The three tanks used here are 20.25 inches x 12.625 inches x 10.5 inches (length x width x height) in size and have a 10-gallon water storage capacity. To get real-time data during experiments, the water level, nitrate, and pH temperature sensors are mounted while the flow sensor is installed on a ½ inch PVC pipe. In addition, a water level sensor is mounted for capturing data from the water reservoir.

### 3.1.2. Bus Topology
In this topology, four rectangular tanks and two diaphragm water pumps are connected via ¾ inch C-PVC pipes, as Figure 8b indicates. In Bus topology, the C-PVC pipe is installed at the centre, alternatively distributing water to two tanks on each side. The water flow also can be changed using manual valves, which are installed for every tank and line. The four tanks in this topology are 16.25 inches x 8.375 inches x 10.5 inches (length x width x height) in size and have a 5.5-gallon water storage capacity. For data collection in real-time, the water level and EC sensors are mounted while the pressure sensor is installed on a main line ¾ inch C-PVC pipe.

### 3.1.3. Star Topology
The Star topology is developed by connecting one medium-size and four small-size water tanks via ½ inch C-PVC pipes. A schematic diagram for Star topology is shown in Figure 8c. This topology is designed using four diaphragm water pumps and five cubical tanks. In this topology, one diaphragm water pump transfers water from the main water reservoir to the central tank (node), and then two additional diaphragm water pumps distribute this water to four other small tanks. The manual valves for each tank are provided to change or control the water flow. Lastly, at the end of the experiment, the fourth pump pulls water from all four tanks back to the reservoir. This process is performed using a four-way water splitter and tubes. The four tubes are immersed in four small tanks to pull the water and transfer it back to the reservoir. The water flow can also be controlled or turned on/off using the individual valves on the splitter. The central tank used in this topology is 15.25 inches x 15.25 inches x 15.25 inches (length x width x height) in size and has a 14-gallon water storage capacity. The four small tanks are 9.25 inches x 9.25 inches x 9.25 inches (length x width x height) in size and each tank has a 3-gallon water storage capacity. For data collection in real-time, water level, pH, temperature, and EC sensors are installed. Also, one water level sensor is mounted to capture data from the reservoir that provides water for this topology.

### 3.1.4 Soil Topology

Vertical farming (Van Gerrewey et al., 2021) involves cultivating plants in multiple layers to maximise yield within a confined space. The definitions of vertical farming differ due to factors such as size, density, control level, layout, building type, location, and purpose. These vertical farming principles are incorporated into the soil topology of ACWA to enhance space efficiency. The ACWA soil topology is mainly dedicated to conducting experiments on agricultural soils and focusing on irrigation and water transportation through them. It is designed in a vertical arrangement, illustrated in Figure 9, (a) shows the representations and (b) shows the real topology.

The vertical arrangement in the Soil Topology is achieved through a 4-Tier metal wire shelving unit, which is 54 inches in height and 36 inches in width. The spacing between each shelf level is 14 inches. This design enables independent experiments on different soil types because each soil pot has its own 3.5L irrigation drip bag and 3-gallon drainage tank. Levels one and three of the shelving unit can accommodate two soil pots each and two water tanks placed on levels two and four. The adjacent tanks are designed to collect drainage from the corresponding soil pots. Further, sensors in each soil pot are mounted to capture moisture, temperature, and EC data. This topology also uses water level and turbidity sensors to capture real-time data from collected drained water.

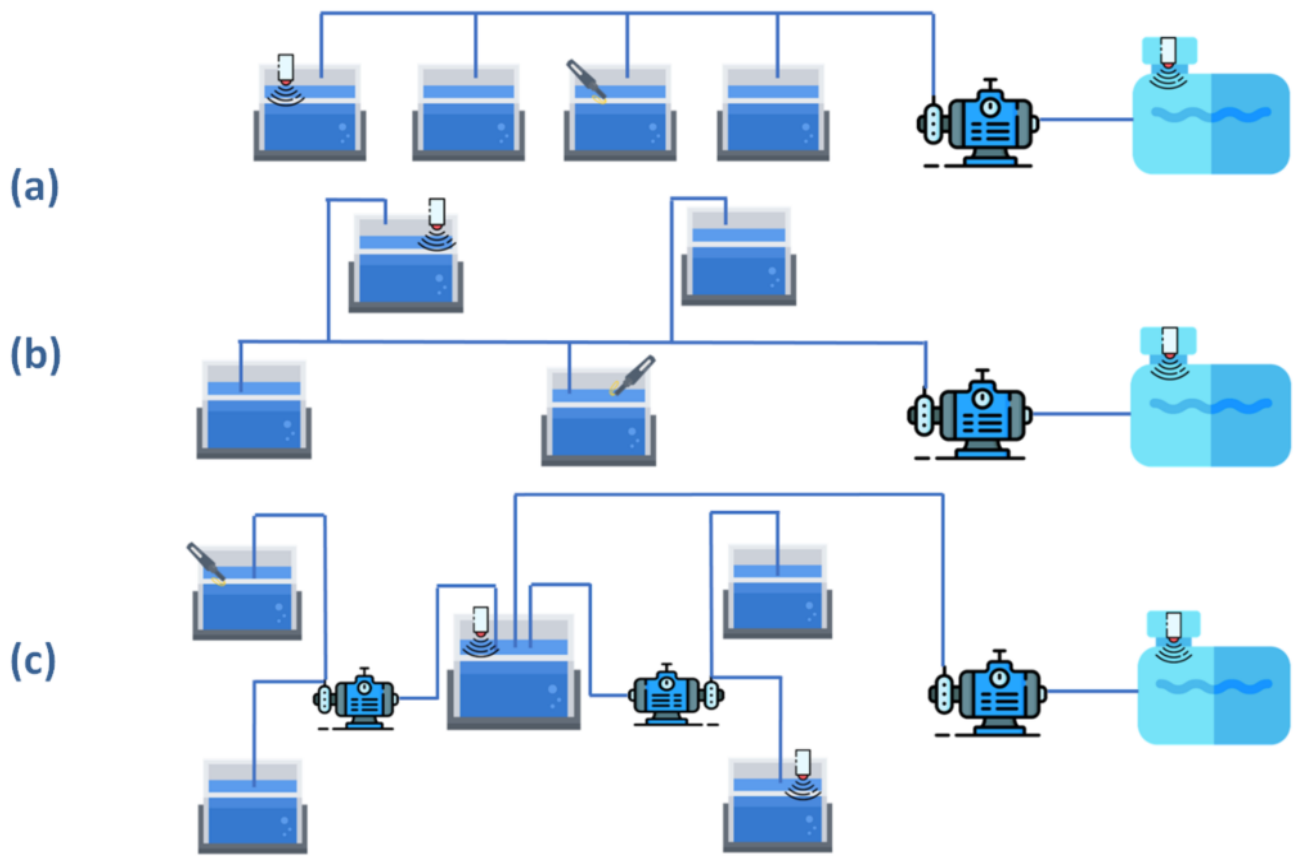

*Figure 8: Schematic representations of the (a) line topology, (b) bus topology, and (c) star topology as used in ACWA*

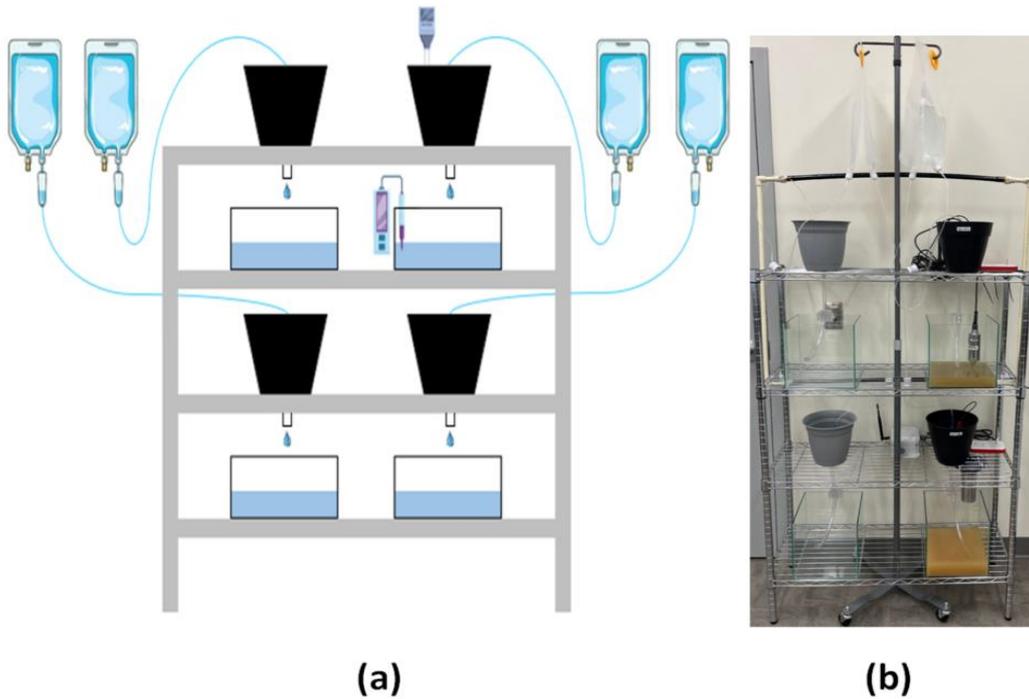

**(a)**                    **(b)**

*Figure 9: (a) Schematic representation of ACWA's soil topology (b) the soil topology*

Associated with Figure 8, Figures 10 and 11 below show the ACWA lab topologies: line and bus (10), and star (11), in Blacksburg, VA, on the Virginia Tech campus.

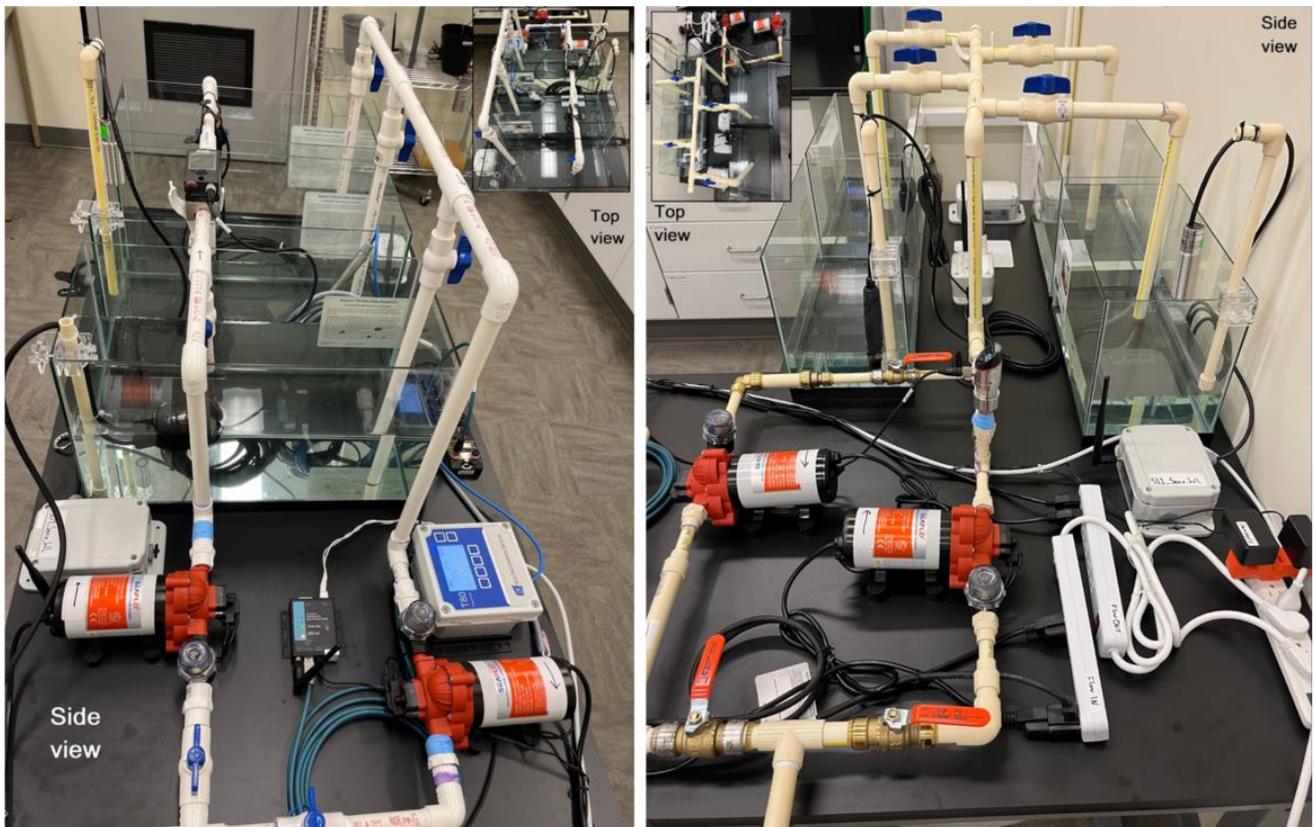

*Figure 10: (left) Line topology, and (right) Bus topology*

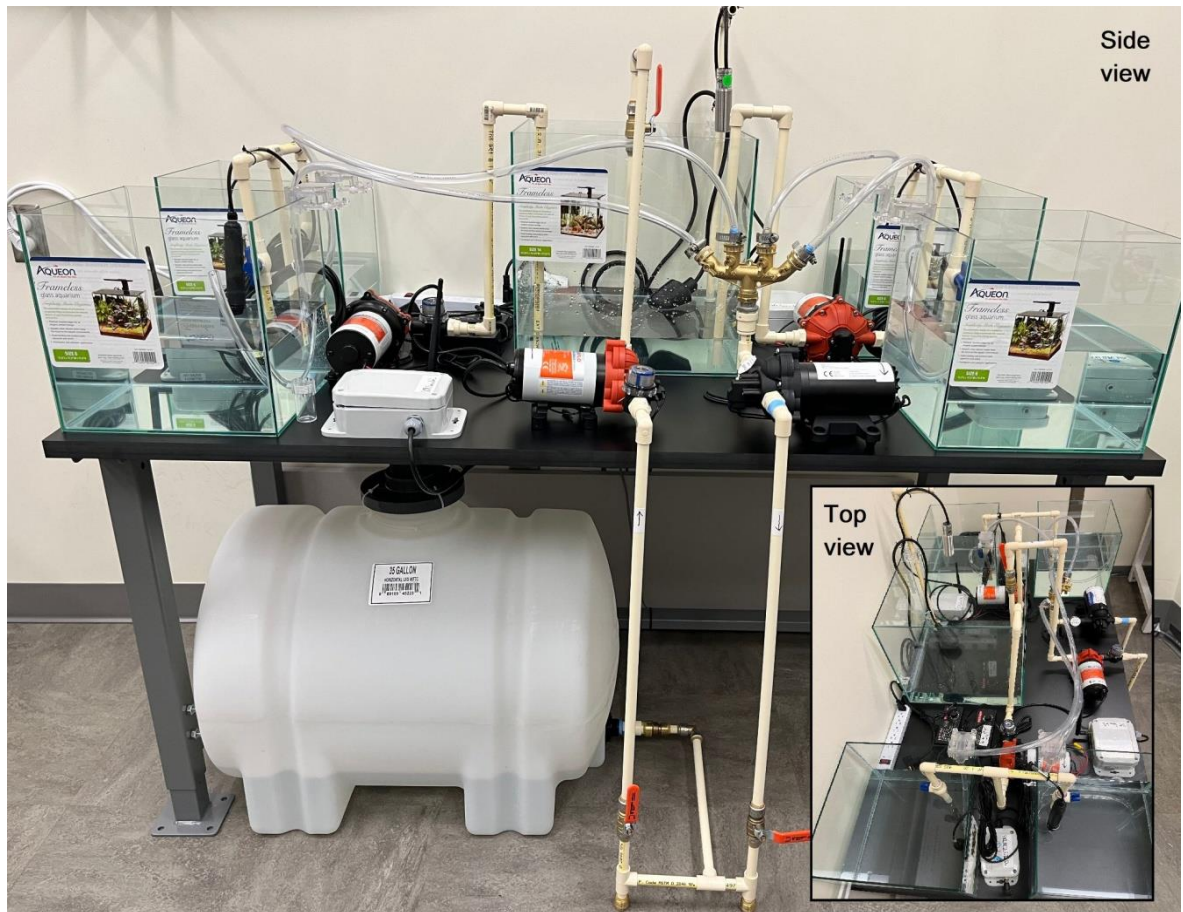

*Figure 11: Star topology, top and side views*

The mentioned 4 topologies are connected to computational nodes and sensors, and via multiple communication protocols; the details of which are presented in the following sections. Appendix A shows other photos from the testbed.

### 3.2 Sensors, Chemicals, and Technologies Deployed

This section includes details on the sensors built into the ACWA topologies, including their communication protocols, as well as the chemicals and solutions applied for AI and Cyber experimentation. The list of sensors is presented in Table 2. Table 3 lists the main experimental chemicals and chemical solutions at ACWA.

*Table 2: Technical details for all ACWA sensors*

| Label | Name | Communication Protocol | Type | Topology |
|-------|------|------------------------|------|----------|
| S01 | S01_NCD_EcTempDo | Zigbee | EC/Temp/DO Sensor | Bus |
| S02 | S02_NCD_EcTempDo | Zigbee | EC/Temp/DO Sensor | Star |
| S03 | S03_NCD_PhTemp | Zigbee | pH/Temp Sensor | Line |
| S04 | S04_NCD_PhTemp | Zigbee | pH/Temp Sensor | Star |
| S05 | S05_NCD_MoistTempEC | Zigbee | Soil Moisture/Temp/EC Sensor | Soil, Pot 1 |

| S06 | S06_Senix_WL | LoRa | Tough Sonic 50 Level Sensor | Line, Bus, Reservoir |
|-----|--------------|------|----------------------------|----------------------|
| S07 | S07_Senix_WL | LoRa | Tough Sonic 50 Level Sensor | Star, Reservoir |
| S08 | S08_Senix_WL | LoRa | Tough Sonic 50 Level Sensor | Soil |
| S09 | S09_Senix_WL | LoRa | Tough Sonic 14 Level Sensor | Line |
| S10 | S010_Senix_WL | LoRa | Tough Sonic 14 Level Sensor | Bus |
| S11 | S011_Senix_WL | LoRa | Tough Sonic 14 Level Sensor | Star |
| S12 | S12_Keyence_Pressure | Modbus | GP-MT | Bus |
| S13 | S13_Keyence_Flow | Modbus | FD-H | Line |
| S14 | S14_Lcom_Turbidity | Modbus | SRWQ100 | Soil, Water Tank |
| S15 | S15_ECD_Nitrate | Modbus | S80 Nitrate Ion Sensor | Line |
| S16 | Soil moisture probe | LoRa | Soilmote | Soil, Pot 1 |
| S17 | Soil moisture probe | LoRa | Soilmote | Soil, Pot 2 |

The sensors are connected to computers and GPU devices via three communication protocols, Modbus, Long Range, and Zigbee, as follows: (1) *Modbus*: A widely used communication protocol (Thomas, 2008) in industrial automation and control systems. It allows different electronic devices, such as sensors, actuators, and PLCs, to exchange data and control commands over a network. Modbus is characterised by its simplicity and versatility, making it suitable for simple and complex applications in various industries. It operates over different physical layers, including serial connections like RS-232, RS-485, and Ethernet, making it adaptable to different communication needs. (2) *Long Range*: LoRa, or Long Range, is a wireless communication technology for long-distance data transmission with low power consumption (Devalal & Karthikeyan, 2018). It enables devices to communicate wirelessly over extended ranges. This makes LoRa useful for applications such as Internet of Things (IoT) devices, smart city solutions, and rural connectivity. LoRa uses spread spectrum modulation techniques to achieve reliable communication even in challenging environments, and its low power requirements make it suitable for battery-operated devices, extending its operational lifespan. (3) *Zigbee*: A wireless communication protocol (Ergen, 2004) developed for creating low-power, short-range networks among various devices. It is designed for home automation and industrial control. The Zigbee protocol enables efficient and reliable data exchange. It operates on the IEEE 802.15.4 standard and uses a mesh network topology. This allows devices to communicate directly or indirectly through other devices, enhancing coverage and reliability. Zigbee's focus on low power consumption makes it well-suited for battery-operated devices in smart homes and IoT applications.

*Table 3: List of chemical solutions in ACWA*

| Type | Concentration(s) | Usage |
|------|------------------|-------|
| pH | 4.01<br>7<br>10.01 | Calibration of NCD pH and temperature sensor for three-point calibration |
| Turbidity | 0 NTU, | Calibration of L-com Turbidity sensor for two- |

| | 100 NTU | point calibration |
|---|---|---|
| EC | 12.88 mS/cm<br>64 mS/cm | Calibration of NCD EC sensor for two-point calibration |
| Nitrate | 10 ppm<br>100 ppm | Calibration of ECD Nitrate sensor for two-point calibration |
| DO | Zero Oxygen | Calibration of NCD DO sensor |
| Sodium Hydroxide | 0.1 M | Experiment design and simulation |
| Distilled Water | --- | To remove mineral buildup from sensors |

To manage data streams for AI development and Cyberbiosecurity measures the following software are used (but not limited to): MongoDB, NodeRed, R Dashboards, Python, TensorFlow, Keras, and other data management technologies, more information on that is listed in the public access GitHub page: https://github.com/AI-VTRC/ACWA-Data. The next section presents the ACWA simulator and its structural details.

### 3.3 The ACWA Simulator

The main goal of the simulation is to showcase water flowing through pipes from a reservoir into two interconnected tanks. In a digital twin environment, a pump is placed between one tank and the reservoir (as Figure 12 illustrates). The pump characteristics will determine the initial flow rate, and a valve can be added between tanks as a control mechanism. The elevation of the bases of the tanks can be changed to simulate gravitational flow. Additionally, hydraulic and water nutrient parameters will be measured throughout the process. These parameters include pH, water level, water pressure, water surface temperature, and changes in nitrate, phosphate, Biochemical Oxygen Demand (BOD), DO, and Sodium Hydroxide (NaOH) concentrations. The connecting pipe diameter and surface roughness can also be programmed to simulate flow through a pipe.

The ACWA simulator generates a CSV file incorporating time series data of all the parameters for every simulation. The novel part of the simulator is that it provides a time series for both hydraulic parameters, such as water level and pressure, along with water nutrient parameters, such as pH, Nitrate, BOD, and NaOH, which change every second. This generated data can be used for calibration, validation, AI modelling, data poisoning experiments, and comparison with the physical system. Inputs to the simulator environment include: pH, BOD, DO, Nitrate, NaOH, Water Level, Water Temperature, Water Density, Kinematic Viscosity, Discharge, Pipe Material, Pipe Length, Pipe Diameter, Tank Shape, Tank Size – the data outcomes of the simulator as presented in Table 4.

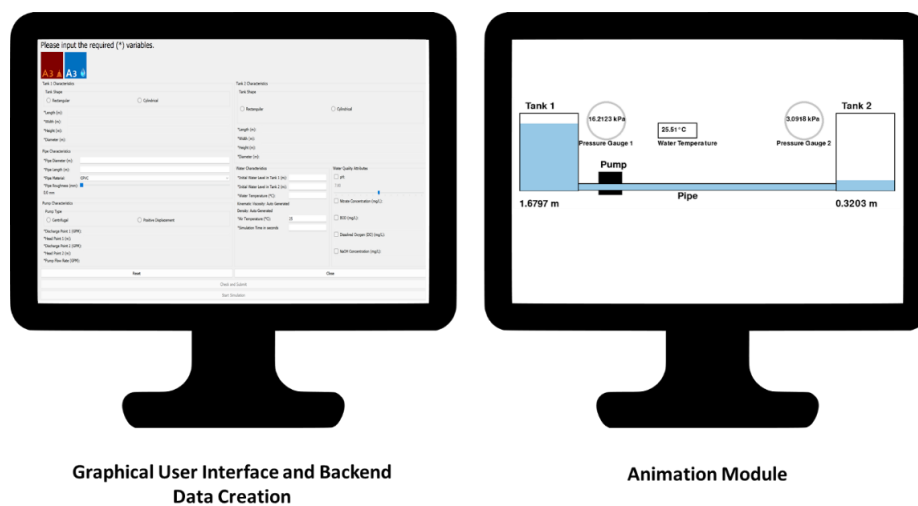

**Graphical User Interface and Backend Data Creation**          **Animation Module**

*Figure 12: Framework of the simulation*

*Table 4: Water quality attributes*

| Column Name | Description | Unit | Data Type | Sample Value |
|---|---|---|---|---|
| Time | Timestep of the simulation | HH:MM:SS | Date/Time | 00:00:00 |
| Reservoir Water Level | Water level in the reservoir | Meters (m) | Float | 9.01 |
| Tank Water Level | Water level in the tank | Meters (m) | Float | 3.07 |
| Pressure at Reservoir Bed | Pressure at the bottom of the reservoir | Pounds per Square Inch (psi) | Float | 18.96 |
| Pressure at Tank Bed | Pressure at the bottom of the tank | Pounds per Square Inch (psi) | Float | 14.68 |
| Water Temperature (°C) | Temperature of the water | Degrees Celsius (°C) | Float | 27.85 |
| pH | pH value of the water | Unitless | Float | 6.49 |
| BOD (mg/L) | BOD level | Milligrams per Litre (mg/L) | Float | 2.88 |
| DO (mg/L) | DO level | Milligrams per Litre (mg/L) | Float | 7.79 |
| Nitrate (mg/L) | Concentration of nitrate | Milligrams per Litre (mg/L) | Float | 16.13 |
| NaOH (mg/L) | Concentration of sodium hydroxide (NaOH) | Milligrams per Litre (mg/L) | Float | 0.78 |

The software simulator has built-in physical assumptions that can be updated and tuned using the code, the following nine assumptions are set:
1. Water is assumed to be an incompressible Newtonian fluid (Mott & Untener, 2015).
2. Water is in steady state i.e., the type of the flow throughout the process will not change.
3. Flow should either be laminar or turbulent. This is implemented in the constraints section.
4. The reservoir and tanks are at atmospheric pressure.
5. Only head loss due to friction, entry, and exit loss is considered.
6. pH interacts with NaOH; BOD interacts with DO. No other mutual interactions are considered. The change in constituents is determined only by wall reaction, bulk reaction, and decay function.
7. Temperature can change only due to differences in air temperature and convection.
8. The model assumes a steady-state scenario, i.e., the conditions are not changing with time, only with position along the pipe. The fluid properties (like density, specific heat, and the heat transfer coefficient) are constant. This might not be the case, especially if the temperature changes significantly.
9. Surface temperature of the pipe is assumed to be constant along the length of the pipe.
Additionally multiple physical characteristics are built into the system, related to the input variables mentioned prior, those are presented in Table 5 (Simulator inputs are presented in Appendix C).

*Table 5: Physical representations in ACWA simulator*

| Input's Physical Representation | Description | Source for Formulas* |
|---|---|---|
| Water Density | Auto-generated and constrained as a function of water temperature. | Kell et al., 1975 |
| Kinematic Viscosity | Expressed as a function of temperature. | Guo, 2020 |
| Overflow Check | Overflow can occur when the volume of water pumped into | Modi, 2019 |

| | a tank exceeds its maximum capacity. Mathematically speaking, the available volume in tank 2 must always be adequate to water transported from tank 1. | |
|---|---|---|
| Pipe Flow Safety Factor | The diameter-to-length ratio is critical to prevent open channel flow and ensure the pipe remains full (avoiding "slug" flow or a mix of air and water). There is no strict ratio for all systems, but it depends on the specifics of the setup; a general guideline for keeping pipes fully pressurised is to ensure sufficient inlet head or pressure. | Mays, 2000 |
| Flow Type | Laminar and turbulent flows are two fundamental flow regimes in fluid dynamics, and they describe how fluid particles move within a conduit - pipe, open channel, or flow medium. Laminar or streamlined flow occurs when fluid flows in parallel layers (or laminae) with minimal mixing or lateral crossover between the layers. Turbulent flow is characterised by chaotic, irregular motion of particles in which fluid particles move randomly and swirl. It is necessary to determine whether the flow is laminar or turbulent. | Mott & Untener, 2015 |
| Elevation Pressure | Conveys how a force is distributed over a particular surface. The water pressure depends on the height of the water column, i.e., how high the water surface is above the tank/reservoir bed. | Mott & Untener, 2015 |
| Energy Loss | The hydraulic movement in the given system is governed by the energy equation derived from Bernoulli's equation of fluid flow. | Mott & Untener, 2015 |
| Pipe Head Loss | When pumped water goes through the pipes and due to friction generated in the pipe, the water will lose energy, resulting in decreased pressure in terms of head loss. | Mott & Untener, 2015 |
| Friction Factor | The Colebrook-White formula can be used to find out the friction factor. | Brandt, 2017 |
| Valves and Joints Loss | Entry and exit loss. | Mott & Untener, 2015 |
| Water Level Rise | Considering all the friction loss and other energy conditions, there will be a specific flow at the end of the pipe and pressure. This will, in turn, transfer into the tank, and the water level will rise depending on the tank area. | Modi, 2019 |
| Advective transport | Longitudinal dispersion is usually not a vital transport mechanism under most operating conditions. This means there is no intermixing of mass between adjacent parcels of water travelling down a pipe. | Chaudhry & Mays, 2012 |
| Mixing in Storage Facilities | It is convenient to assume that the contents of storage facilities (tanks and reservoirs) are entirely mixed. This is a reasonable assumption for many tanks operating under fill-and-draw conditions, providing sufficient momentum flux is imparted to the inflow. | Rossman & Grayman, 1999 |
| Heat Transfer | For heat transfer of the fluid through pipe, convection theorem is used to model convective heat transfer from the outside environment to the water through the pipe. | Bergman, 2011 |

*All formulas are presented in Appendix B.*

### 3.4 ACWA Data

The data created from all the mentioned parts in this section are the main outcome of ACWA; the data variables are presented in Table 6 below (data samples are available in the mentioned public GitHub repository; other datasets are available upon request from the authors).

*Table 6: Data descriptions for all ACWA sensor variables*

| EC Temperature DO Sensor (S01, S02) | | |
|---|---|---|
| Field Value | Description | Example Value |
| addr | Device Address | 00:13:a2:00:42:29:e7:3d |
| battery | Battery voltage | 3.29 |
| battery_percent | Battery life | 99.64 |
| counter | Counter | 113 |
| firmware | Firmware version | 2 |
| nodeId | Node ID | 0 |
| original.data | Raw data | [127,0,2,3,255,113,0,66,0,0,0,0,0,0,0,0,0,0,8,121,0,0,1,213,0,0,20,243,8,194] |
| original.mac | Mac Address | 00:13:a2:00:42:29:e7:3d |
| original.rssi | Raw RF signal strength | [object Object] |
| original.type | Type | receive_packet |
| received | Data Received size | 1.68E+12 |
| rssi | RF signal strength | 40 |
| sensor_data.DO | DO | 4.69 |
| sensor_data.DO_Saturation | DO saturation | 53.63 |
| sensor_data.EC | EC | 0 |
| sensor_data.Salinity | Salinity | 0 |
| sensor_data.TDS | TDS | 0 |
| sensor_data.Temp | Temperature | 21.69 |
| sensor_data.Temp_DO | Temperature of DO | 22.42 |
| sensor_name | Name | EC and DO and Temperature Sensor |
| sensor_type | Sensor Type | 66 |
| type | Type | sensor_data |
| pH Temperature Sensor (S03, S04) | | |
| Field value | Description | Example Value |
| addr | Device Address | 00:13:a2:00:42:29:e7:06 |
| battery | Battery voltage | 3.29 |
| battery_percent | Battery life | 99.64 |
| counter | Counter | 1 |
| firmware | Firmware version | 4 |
| nodeId | Node ID | 0 |
| original.data | Raw data | [127,0,4,3,255,1,0,61,0,3,120,8,85] |
| original.mac | Mac Address | 00:13:a2:00:42:29:e7:06 |

| original.receive_options | | |
|---|---|---|
| original.rssi | Raw RF signal strength | [object Object] |
| original.type | Type | receive_packet |
| received | Data Received size | 1.68E+12 |
| rssi | RF signal strength | 40 |
| sensor_data.Temp | Temperature | 21.33 |
| sensor_data.pH | pH | 8.88 |
| sensor_name | Name | pH and Temperature Sensor |
| sensor_type | Sensor Type | 61 |
| type | Type | sensor_data |
| **Soil Moisture Temperature EC Sensor (S05)** | | |
| Field value | Description | Example Value |
| addr | Device Address | 00:13:a2:00:42:29:e7:43 |
| battery | Battery voltage | 3.29 |
| battery_percent | Battery life | 99.64 |
| counter | Counter | 123 |
| firmware | Firmware version | 4 |
| nodeId | Node ID | 0 |
| original.data | Raw data | [127,0,4,3,255,123,0,69,0,0,0,0,0,0,8,122,0,0,0,0,0,0,0,0] |
| original.mac | Mac Address | 00:13:a2:00:42:29:e7:43 |
| original.receive_options | | |
| original.rssi | Raw RF signal strength | [object Object] |
| original.type | Type | receive_packet |
| received | Data Received size | 1.68E+12 |
| rssi | RF signal strength | 40 |
| sensor_data.EC | EC | 0 |
| sensor_data.Moisture | Moisture | 0 |
| sensor_data.Temperature | Temperature | 21.7 |
| sensor_name | Name | Soil Moisture Temperature EC Sensor |
| sensor_type | Sensor Type | 69 |
| type | Type | sensor_data |
| **Water Level Sensor (S06, S07, S08, S09, S10, S11)** | | |
| Field value | Description | Example Value |
| Counts | Sensor counts in the latest cycle | 20365 |
| DI | | 1 |
| Distance | Target distance in mm | 275.660634 |
| FWVer | Transmitter firmware version | 1.12 |
| FaultMsg | Displayed message if any | |
| Flags | | 0 |
| FriendlyName | Displayed sensor name | AirWire 1005 |
| Hyst | Hysteresis in units, alarm to reset | 0 |
| LOffset | Linear Offset in units, number line zero (*) | 0 |
| LScale | Scale, Units to sensor Counts | 0.013536 |
| LUnits | Linear Units displayed as Tag | Inches |
| Lat | Latitude of sensor (not used) | 44.3315392 |
| LifeCount | Lifetime count of transmitter | 29993 |
| Log | Log text displayed | AirWire 1005 measured 275.66 Inches on May 3 @ 14:08:03 |
| LogSpace | | 7373 |
| Lon | Longitude of sensor (not used) | -73.111984 |
| OnTime | Transmitter time at full power, D-HH:MM: SS | 1-06:01:11 |
| RSSI | RF signal strength (**) | -59 |
| RangeMax | Max range displayed on Bar Graph | 600 |
| RangeMin | Min range displayed on Bar Graph | 0 |
| RateTime1 | Start SampleRate1 (24 hr. mode) | 8:00 |
| RateTime2 | Start SampleRate2 (24 hr. mode) | 17:00 |
| RawCounts | Sensor counts in the latest cycle (unfiltered) (***) | 20593 |
| RetryMax | Maximum transmit retries | 0 |
| SNR | - | 12.2 |
| SampleRate1 | Sample Rate 1 in seconds | 60 |
| SampleRate2 | Sample Rate 2 in seconds | 60 |
| SensorFWVer | Sensor firmware version | 47 |
| SensorHardwareId | Sensor hardware identification | 1017 |
| SensorProductId | Sensor model identification | 4006 |
| SensorProductName | Sensor brand name | ToughSonic 50P |
| SensorTemp | Sensor internal temperature (deg F) | 68.44 |
| ShotCount | Measure cycles since last reset | 29993 |
| Time | Date and Time of latest data | 08:03.2 |
| TransmitterBat | Voltage level, transmitter (v) | 3.657 |
| TransmitterTemp | Temperature, transmitter (deg F) | 60.6 |
| TxFaultCount | Sum of transmit retries since reset | 2632 |
| UTCmsec | Universal time from January 1, 1970 (msec) | 1.68E+12 |
| _id | - | 6452a304f5626fa35f3c08a5 |
| alarm | Current alarm displayed | - |
| cmd | - | 64 |
| ecode | - | 20 |
| eui | End Unit Identifier of transmitter | 00-80-00-00-00-01-99-bf |
| filename | Path and filename of current active log in receiver gateway | /media/card/senix/00-80-00-00-00-01-99-bf/AirWire_00-80-00-00-00-01-99-bf_2023.05.03 |
| gweui | End Unit Identifier of gateway | 00-80-00-00-a0-00-6d-0d |
| icon | - | wifi |

| time | Date and time of latest data | 2023-05-03T18:08:03.175498Z |
| timetext | Timestamp on display | May 3 @ 14:08:03 |
| tstr115 | - | Critical H |
| tstr67 | - | AirWire 10 |

| **Pressure Sensor (S12)** | | |
| --- | --- | --- |
| Field value | Description | Example Value |
| pressure | Pressure | 0 |
| pressure_unit | Pressure unit | psi |
| temperature | Temperature | 24.2 |
| temperature_unit | Temperature unit | celcius |
| timestamp | Timestamp | 1.68317E+12 |

| **Flow sensor (S13)** | | |
| --- | --- | --- |
| Field value | Description | Example Value |
| flow | Flow rate | 0 |
| flow_acc | Flow accumulation | 7.3 |
| flow_acc_unit | Flow accumulation unit | gallon |
| flow_unit | Flow rate unit | g/min |
| temperature | Temperature | 31.8 |
| temperature_unit | Temperature unit | celcius |
| timestamp | Timestamp | 1.68317E+12 |

| **Lcom - Turbidity sensor (S14)** | | |
| --- | --- | --- |
| Field value | Description | Example Value |
| current | Output (4mA-20mA) | 16.875 |
| current_unit | Output unit | mA |
| temperature | Temperature | 25.109993 |
| temperature_unit | Temperature unit | celcius |
| timestamp | Timestamp | 1.68317E+12 |
| turbidity | Turbidity | 3232 |
| turbidity_unit | Turbidity unit | NTU |

| **Nitrate Ion sensor (S15)** | | |
| --- | --- | --- |
| Field value | Description | Example Value |
| current | Output (4mA-20mA) | 16.875 |
| current_unit | Output unit | mA |
| temperature | Temperature | 25.109993 |
| temperature_unit | Temperature unit | celcius |
| timestamp | Timestamp | 1.68317E+12 |
| nitrate_ion | Nitrate Ion | 45.14 |
| nitrate_ion_unit | Nitrate unit | ppm |

| **Soil Probes (S16 and S17)** | | |
| --- | --- | --- |
| Field value | Description | Example Value |
| mostRecentData.soilmoisture.t | Time of day | 2023-09-16T17:30:00+00:00 |
| mostRecentData.soilmoisture.u | Moisture percentage | % |
| mostRecentData.soilmoisture.v | Moisture value | 6.66 |

| mostRecentData.soilmoisture.i | Index | 5 |
|---|---|---|
| mostRecentData.soilmoisture.via reportingInterval | data collection frequency | 5 |
| reportingVia | Unit | mqtt |

Additionally, data collected via the simulators includes the following variables: Time (seconds), Tank 1 Water Level (m), Tank 2 Water Level (m), Tank 1 Pressure (Pa), Tank 2 Pressure (Pa), Nitrate Concentration (mg/L), BOD Concentration (mg/L), DO Concentration (mg/L), pH, Temperature (Degree Celsius). All data are visualised in multiple data dashboards (presented in Appendix D).

## 4. Results and Discussion

The scope of the ACWA lab is extensive, covering a diverse range of applications relevant to its design and operation. It includes various concepts, such as data creation and collection, benchmark scenarios for Cyberbiosecurity research (Sobien et al., 2023), AI assurance techniques to address adversarial attacks, fine-tuning process of soft sensors, innovative ideas for water and agricultural-related public policies, and more. Considering these aspects, the potential use cases that can be developed in the ACWA lab include (but not limited to):

1. Digital twin: A digital twin is a replication of a physical sensor that allows monitoring, visualisation and prediction of its physical counterpart (Eckhart & Ekelhart, 2019). The data collected from water and soil sensors could be used for development of digital twins (besides the presented ACWA simulator). This can help in different applications such as soil quality, systems' simulation, along with other potential agriculture and water-related uses (Eckhart & Ekelhart, 2018).

2. Data poisoning: The EPA (Environmental Protection Agency) has established water regulation limits, and water distribution plants use sensors to control the level of chemicals (such as Phosphorus) in the water. The ACWA lab allows the simulation of data poisoning via advanced AI algorithms that can help simulate cases where the data used for decision-making are compromised, similar to the many cyber incidents presented by Hassanzadeh et al., 2020.

3. Water quality: WSSs uses filtration and chlorination processes to remove turbidity from water (Stevenson & Bravo, 2019). The use of high chlorination forms trihalomethanes (THMs) and Haloacetic acids (HAAs), which can create human health hazards (Lowe et al., 2022). In these scenarios, AI can be used to develop an early warning system for predicting reclaimed water (U.S. Environmental Protection Agency, 2023) turbidity. This may help WSS in planning the use of the chlorination process efficiently.

4. Pump operations: These operations, such as determining capacity, number of pumps, and when to start them, can be determined optimally. This can assist pump operators by providing actionable recommendations on pump operations and reduce pump usage and energy consumption.

5. Simulation: The ACWA simulator generates different scenarios to showcase water flow from a reservoir to a tank. The simulator provides complete control of inputs such as selecting material for the pipe, tank dimensions, and chemicals. This simulator can be used to study the effects of different conditions (chemicals, pipe materials, etc.) on the water.

6. Decision-support system: A decision-support system represents the role of technology, such as AI, to provide the stakeholders with intuition, insight, and understanding (Keen, 1980; Brill et al., 1990). These systems can be developed by combining advanced AI techniques with AI assurance for different water and agricultural stakeholders.

7.  Process monitoring: ACWA data visualisations can represent the information and act as a diagnostic tool to bring situational awareness (Few, 2007). This can help monitor different agricultural and/or water plant processes in real-time to check for subtle accuracies ingested into sensor networks.

8.  Data generation: The need for more datasets for securing Cyber-Physical Systems (CPS) (Goh et al., 2017) can be tackled by generating data from the ACWA lab topologies. ACWA lab provides a facility to generate realistic datasets for network, physical properties, and soil and water nutrient data with sufficient complexities of WSS, a notion not found in any other testbed.

9.  Fertigation: Fertigation is injecting fertilisers and other soil products into the soil (Raut et al., 2017). ACWA lab allows experimentation of different fertiliser concentrations along with other parameters such as pH, and moisture content. This can be useful for instance for developing a recommendation system using AI models for farmers.

10. Economics and policy making: The use of AI, specifically Deep and Reinforcement Learning (DL and RL) - subfields of AI - has been increasingly assisting farm, forest, and ranch managers in decision-making (National, n.d.). In ACWA, advanced AI methods with causal aspects will be researched focusing on applications such as water quality, agricultural market structures, agricultural production, resource allocation, and international trade.

11. Anomaly detection: It is common in smart farms to have physical devices such as sensors, actuators, network cables, and routers outdoors. Also, these devices do not have tamper-resistant boxes, and the changes of intentional or unintentional physical modifications of these devices can occur (Rettore de Araujo Zanella et al., 2020). These anomalies can be identified using AI, and data generated from these devices can be detected in real time, a cyber scenario that can be tested at ACWA.

12. Network attack detection: A report published by the US Council of Economic Advisors (2018) reported >11 large scale cyber incidents in the food and agriculture sector in 2016 for instance. This creates a need to provide robust security solutions, mainly when digital systems and IoT-based technologies are heavily used in agriculture and water domains. The advanced sensors used in ACWA lab can help create various anomalous datasets with a combination of different network attacks such as Denial of Service (DOS), man-in-the-middle, and Address Resolution Protocol (ARP) Spoofing that may help in the development of assured AI models for network attacks' mitigation and detection.

13. Soft sensors: In various industrial and scientific applications, accurate and reliable sensor measurements are crucial for making informed decisions and maintaining high-quality processes (Qin et al., 2000). However, hardware sensors are subject to environmental conditions, calibration drift uncertainties, expensive prices, hard set up, and other unreliable aspects (Geng et al., 2015). One solution to this problem is the development of soft sensors, which are mathematical models that output specific target parameters using hardware sensor data (Ching et al., 2021). This solution can be explored in ACWA considering the availability of advanced high-quality water and soil sensors.

Other novel use cases will be researched considering the need and advancement of water technologies. Additionally, the ACWA lab actively collaborates with external organisations, industry partners, and academic institutions, fostering a collaborative community dedicated to addressing other potential complex challenges in water and agriculture.

## 5. Conclusion

This manuscript presents the new ACWA testbed, a novel cyber-physical system designed and developed to test, simulate, and experiment with cutting edge technologies such as AI and cyber solutions for intelligent water systems. As established in the literature review section, ACWA is the first of its kind in the domain, and it fills a needed gap for evaluating existing challenges and validating some of the proposed solutions in academia and industry. It is rather difficult to represent all potential AI applications in the water sector, for that reason, the four ACWA topologies (line, bus, star, and soil) are developed in a modular manner that allows researchers to manipulate the structure, expand on it, and collect different forms of datasets based on the desired outcomes.

**Data Availability Statement:** ACWA data samples are available in our open access GitHub repository (https://github.com/AI-VTRC/ACWA-Data). More detailed datasets including requests for specific simulations and complete timeseries are available upon request from the authors.

**APPENDICES**

**APPENDIX A:** Other Lab Figures

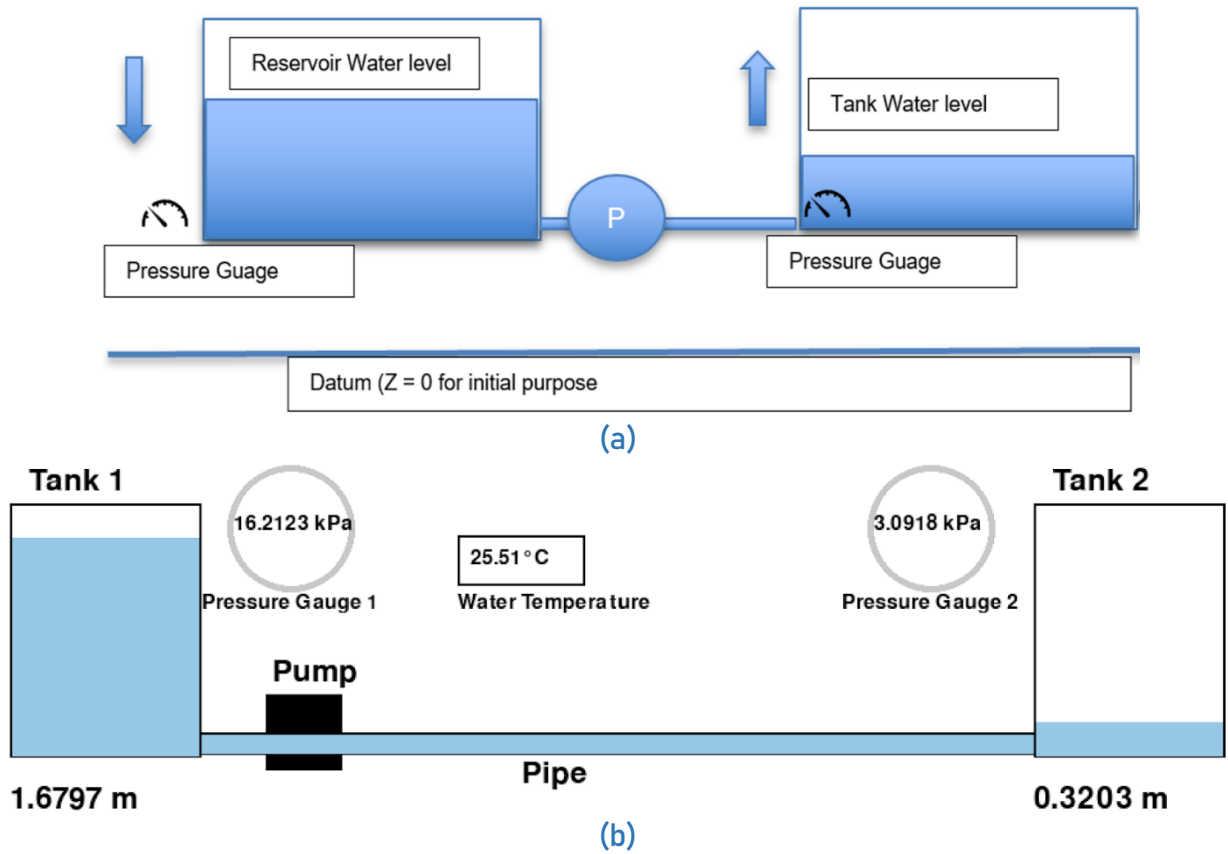

(a)

(b)

*Figure A.1: Simulator's (a) logical design and (b) viz interface*

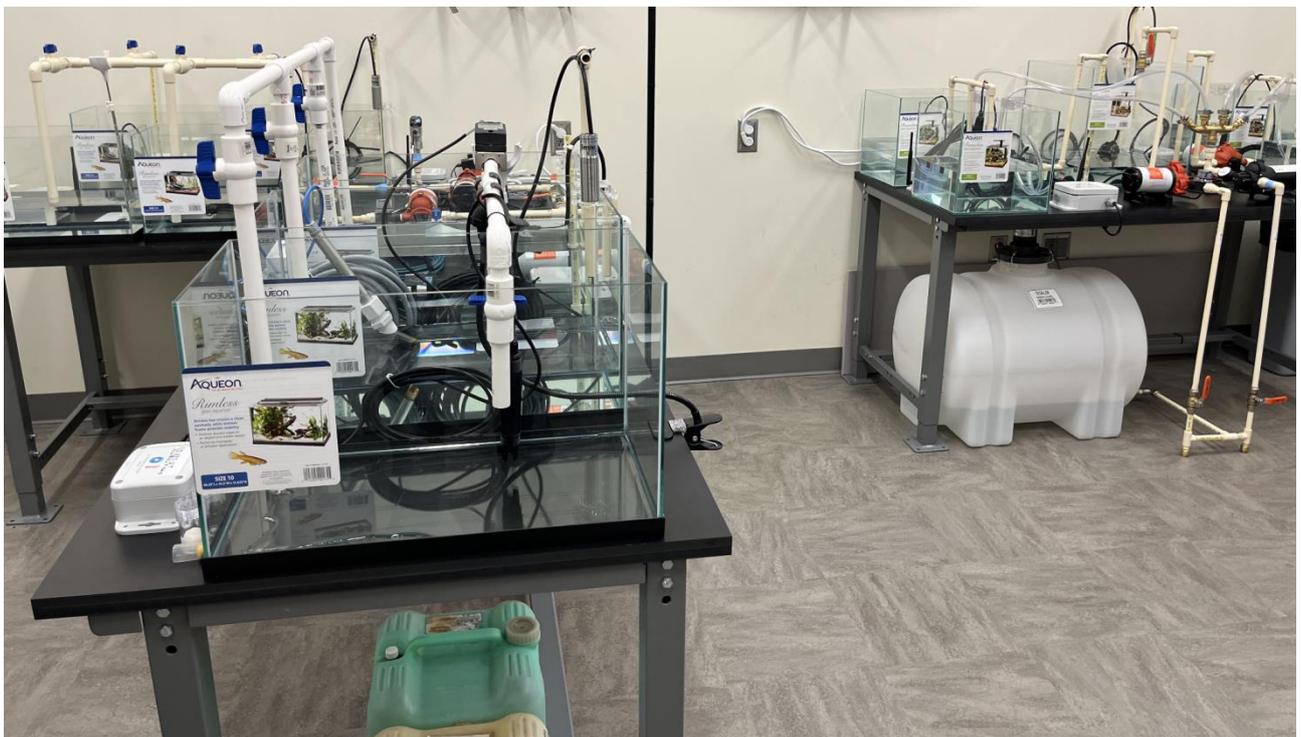

*Figure A.2: ACWA Lab's 3 Water Topologies*

**APPENDIX B:** Formulas for the ACWA Simulator

This appendix presents the formulas referenced in the simulator section of this manuscript.

B.1 Kinematic viscosity:

$$K = \frac{1.773 \times 10^{-3}(1 + 0.0337t + 0.00022t^2)^{-1}}{999.457(1 + 0.000052939t - 0.0000065322t^2 + 0.00000001445t^3)}$$

B.2 Overflow check:

$$A_{tank2} \times (h_{tank2} - WL_{tank2}) \geq A_{tank1} \times WL_{tank1}$$

B.3 Pipe flow:

$$WL_{tank1} > 1.5D$$

B.4 Flow type:

$$N_R = \frac{v \times D}{\nu}$$

B.5 Energy equation:

$$\frac{p}{\gamma} + z + \frac{v^2}{2g} = Constant$$

B.6 Pipe head loss:

$$h_L = f \times \frac{L}{D} \times \frac{v^2}{2g}$$

B.7 Laminar flow:

$$f = \frac{64}{N_R}$$

B.8 Turbulent flow:

$$\frac{1}{\sqrt{f}} = -2log_{10}\left\{\frac{\epsilon}{3.71D} + \frac{2.51}{N_R\sqrt{f}}\right\}$$

B.9 Joint losses:

$$h_{ml} = K \times \frac{v^2}{2g}$$

B.10 Heat transfer:

$$T = T_s - (T_s - T_{in})exp(-\frac{hP}{C_p\hat{m}}z)$$

**APPENDIX C:** ACWA Simulator Input File Data

This shows the list of inputs (with examples) that control any experiment in the ACWA digital twin; the outputs are presented in section 3.4.

"Tank 1 Type": "Rectangular",

    "Tank 1 Length": "0.5",
    "Tank 1 Width": "0.3",
    "Tank 1 Height": "0.3",
    "Tank 1 Diameter": "",
    "Tank 2 Type": "Rectangular",
    "Tank 2 Length": "0.5",
    "Tank 2 Width": "0.3",
    "Tank 2 Height": "0.3",
    "Tank 2 Diameter": "",
    "Pipe Diameter": "0.1",
    "Pipe Length": "3",
    "Pipe Material": "PVC",
    "Pipe Roughness": 0.02,
    "Pump Type": "Positive Displacement",
    "Tank 1 Initial Water Level": "0.2",
    "Tank 2 Initial Water Level": "0",
    "Water Temperature": "26",
    "Air Temperature": "20",
    "Kinematic Viscosity": 0.63,
    "Density": 981.8,
    "Water pH": "7.00",
    "Nitrate Concentration": "10",
    "BOD": "",
    "DO": "",
    "NaOH Concentration": "",
    "Simulation Time": "300",
    "Pump Flow Rate": "3.5",
    "Unique ID": "20230916205552_1A67EF"

**APPENDIX D:** Data Dashboards

This appendix shows the 3 user interfaces for observing experiments in ACWA.

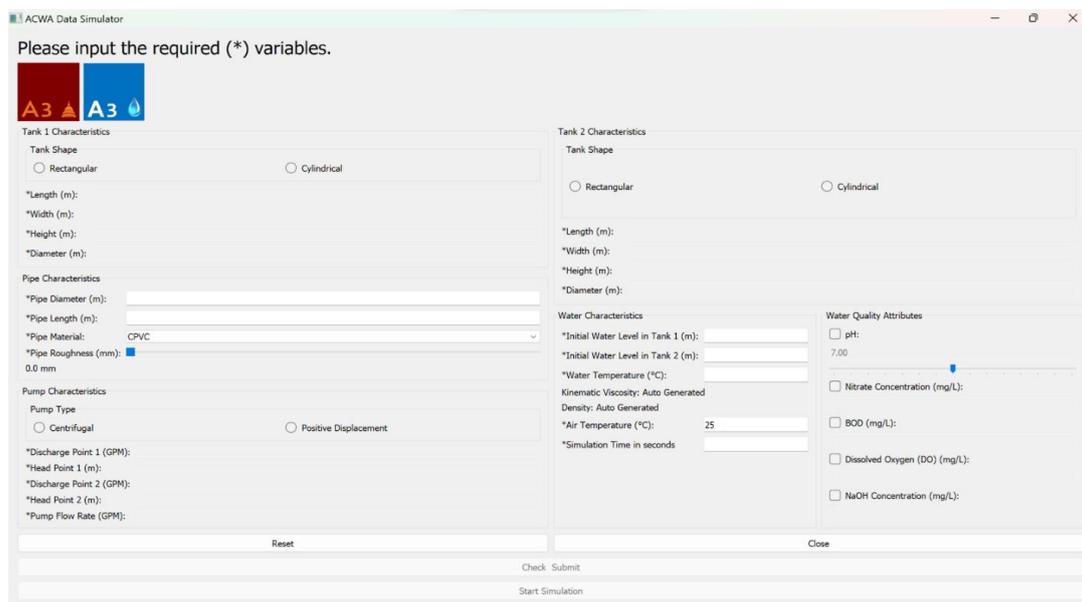

*Figure D.1: ACWA simulator's python interface*

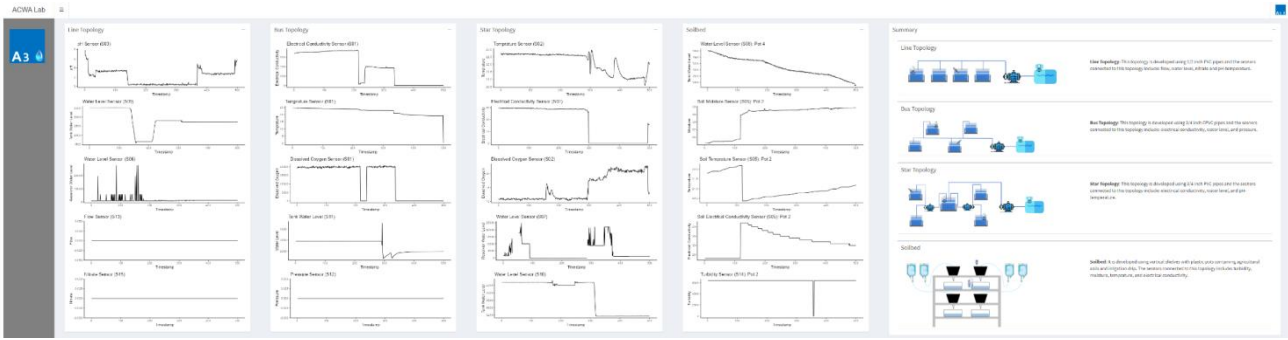

*Figure D.2: ACWA lab dashboard provides real-time updates in the system*

**APPENDIX E:** EPANET Graph and Information

The ACWA testbed has also been designed using EPANET shown in Figure D.1.

EPANET is a public domain water distribution and modelling software package developed by the United States Environmental Protection Agency's (USEPA) Water Supply and Water Resources Division. EPANET can be utilized for hydraulic simulations and studying water quality behaviour within pressurized pipe networks (Rossman, 2000).

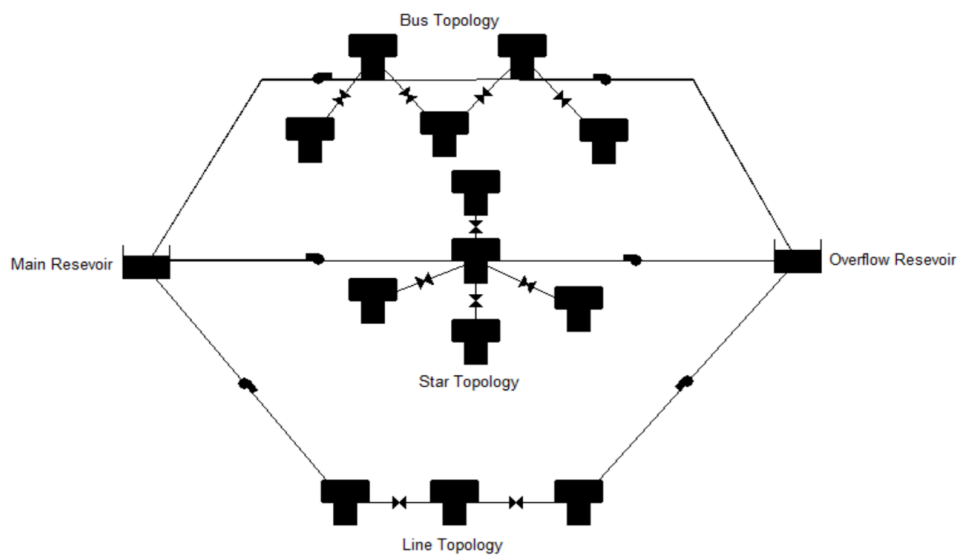

*Figure E.1: The ACWA testbed EPANET network.*

The ACWA testbed EPANET network, the main reservoir acts as a water source, while the overflow reservoir stores the water from all the topologies. It includes three topologies, as discussed in Section lll**,** which receive water from the same source in this simulation. The design decisions in this network are inspired by the design of the ACWA lab. Thus, the network's arrangement and number of components (pumps, pipes, valves, and tanks) are based on the ACWA lab design. Additionally, it offers more than 45 EPANET benchmark networks of various sizes (small, medium, and large) for experimentation.